\documentclass[conference]{IEEEtran}
\usepackage{cite}
\usepackage{amsmath,amssymb,amsfonts}
\usepackage{algorithmic}
\usepackage{graphicx}
\usepackage{textcomp}
\usepackage{xcolor}
\usepackage{glossaries}
\usepackage{siunitx}
\usepackage{hyperref}
\usepackage{cleveref}
\usepackage{booktabs}
\usepackage{makecell}
\usepackage{subcaption}
\usepackage{balance}
\def\BibTeX{{\rm B\kern-.05em{\sc i\kern-.025em b}\kern-.08em
    T\kern-.1667em\lower.7ex\hbox{E}\kern-.125emX}}
\newacronym{nic}{NIC}{Network interface card}
\newacronym{iiot}{IIoT}{Industrial Internet of Things}
\newacronym{opc}{OPC}{Open Platform Communications}
\newacronym{tas}{TAS}{Time-Aware Shaper}
\newacronym{cbs}{CBS}{Credit-Based Shaper}
\newacronym{tsn}{TSN}{Time Sensitive Networking}
\newacronym{taprio}{TAPRIO}{Time-Aware Priority Shaper}
\newacronym{etf}{ETF}{Earliest Time First}
\newacronym{mqprio}{MQPRIO}{Multiqueue Priority}
\newacronym{cots}{COTS}{commercial off-the-shelf}
\newacronym{rtt}{RTT}{round trip time}
\newacronym{ua}{UA}{Unified Architecture}
\newacronym{opcua}{OPC UA}{OPC Unified Architecture}
\newacronym{p2p}{P2P}{point-to-point}
\newacronym{ptp}{PTP}{Precision Time Protocol}
\newacronym{avb}{AVB}{Audio Video Bridging}
\newacronym{gm}{GM}{Grandmaster Clock}
\newacronym{pubsub}{PubSub}{publish-subscribe}
\newacronym{tc}{tc}{traffic control}
\newacronym{sr}{SR}{stream reservation}
\newacronym{qdisc}{qdisc}{queuing discipline}
\newacronym{ecdf}{ECDF}{Empirical Cumulative Distribution Function}
\newacronym{uadp}{UADP}{Unified Architecture Datagram Protocol}
\newacronym{be}{BE}{best-effort}
\newacronym{kpi}{KPI}{key performance indicator}
\newacronym{pcp}{PCP}{Priority Code Point}
\newacronym{skb}{SKB}{socket buffer}
\newacronym{fqcodel}{FQ\_CoDel}{Fair Queuing with Controlled Delay}
\newacronym{udp}{UDP}{User Datagram Protocol}
\newacronym{mqtt}{MQTT}{Message Queuing Telemetry Transport}
\newacronym{mac}{MAC}{Media Access Control}
\newacronym{ip}{IP}{Internet Protocol}
\newacronym{qos}{QoS}{Quality of Service}
\newacronym{ws}{WS}{window size}
\newacronym{cpu}{CPU}{central processing unit}
\newacronym{ram}{RAM}{Random-Access Memory}
\newacronym{is}{IS}{Interframe Spacing}

\DeclareSIUnit[per-mode=symbol]\bps{\bit\per\second}
\DeclareSIUnit[per-mode=symbol]\kbps{\kilo\bps}
\DeclareSIUnit[per-mode=symbol]\Mbps{\mega\bps}
\DeclareSIUnit[per-mode=symbol]\Gbps{\giga\bps}
\DeclareSIUnit[per-mode=symbol]\nanosec{\nano\second}
\DeclareSIUnit\microsec{\SIUnitSymbolMicro s}
\DeclareSIUnit\byte{B}
\DeclareSIUnit\bit{bit}
\DeclareSIUnit\terabyte{TB}
\begin{document}

\title{Real-Time Performance of OPC UA}

\makeatletter
\newcommand{\linebreakand}{
  \end{@IEEEauthorhalign}
  \hfill\mbox{}\par
  \mbox{}\hfill\begin{@IEEEauthorhalign}
}
\makeatother

\author{\IEEEauthorblockN{Erkin Kirdan}
\IEEEauthorblockA{\textit{Framatome} \\
erkin.kirdan@framatome.com}
\and
\IEEEauthorblockN{Filip Rezabek}
\IEEEauthorblockA{\textit{Technical University of Munich} \\
filip.rezabek@tum.de}
\and
\IEEEauthorblockN{Nikolas Mühlbauer}
\IEEEauthorblockA{\textit{Technical University of Munich} \\
n.muehlbauer@tum.de}
\linebreakand
\IEEEauthorblockN{Georg Carle}
\IEEEauthorblockA{\textit{Technical University of Munich} \\
carle@tum.de}
\and
\IEEEauthorblockN{Marc-Oliver Pahl}
\IEEEauthorblockA{\textit{IMT Atlantique} \\
marc-oliver.pahl@imt-atlantique.fr}
}

\maketitle

\begin{abstract}
OPC UA is an industry-standard machine-to-machine communication protocol in the Industrial Internet of Things.
It relies on time-sensitive networking to meet the real-time requirements of various applications.
Time-sensitive networking is implemented through various queueing disciplines (qdiscs), including Time Aware Priority, Multiqueue Priority, Earliest TxTime First, and Credit-Based Shaper.
Despite their significance, prior studies on these qdiscs have been limited to a few.
They have often been confined to point-to-point network topologies using proprietary software or specialized hardware.
This study builds upon existing research by evaluating all these qdiscs in point-to-point and bridged topologies using open-source software on commercial off-the-shelf hardware.
We first identify the optimal configuration for each qdisc and then compare their jitter, latency, and reliability through experiments.
Our results show that open-source OPC UA on commercial off-the-shelf hardware can effectively meet the stringent real-time requirements of many industrial applications and provide a foundation for future research and practical deployments.
\end{abstract}


\section{Introduction}
\label{sec:introduction}
\gls{opc} \gls{ua} is a machine-to-machine communication protocol for industrial automation developed by the \gls{opc} Foundation \cite{IEC62541}. With its platform-independent design, \gls{opc} \gls{ua} offers an efficient and secure framework for interoperability between different systems and devices. It supports complex data types and object models, making it versatile for various industrial applications. \gls{opc} \gls{ua} ensures security measures, including encryption, authentication, and authorization, to safeguard against unauthorized access and cyber threats in critical infrastructures. Its ability to integrate with different hardware and software and its support for client-server and publisher-subscriber communication models makes \gls{opc} \gls{ua} a key enabler for the Industrial Internet of Things.

By integrating \gls{tsn}, \gls{opc} \gls{ua} meets industrial processes' stringent timing and reliability requirements, ensuring synchronized and timely communication between devices and systems. This integration is pivotal for supporting real-time operations in Industry 4.0, where precise timing and coordination across diverse components are essential.

\gls{tsn} is a set of IEEE standards developed to improve the reliability and determinism of standard Ethernet networks. It enables the precise timing and synchronization of data packets across a network, ensuring low latency and minimal jitter. \gls{tsn} incorporates features like time synchronization, traffic scheduling, and resource reservation, allowing for the coexistence of regular and time-critical traffic on the same network.

Despite the crucial role of real-time communication in \gls{opc} \gls{ua} field devices \cite{8610105}, there needs to be more research that explores the capabilities of open-source \gls{opc} \gls{ua} on \gls{cots} hardware.
The current studies primarily rely on specialized hardware and propriety software.
They examine a limited number of \glspl{qdisc}, preventing a comprehensive comparison and identification of the optimal configuration for various requirements.

Our research aims to answer the following questions:

\textbf{Q1:} How does open-source \gls{opc} \gls{ua} perform on \gls{cots} hardware?

\textbf{Q2:} How do various \glspl{qdisc} affect performance?

\textbf{Q3:} How does the presence of a bridge and cross-traffic affect performance?

\textbf{Q4:} What is the optimal configuration for each \gls{qdisc}?

To determine the real-time performance of \gls{opc} \gls{ua} over \gls{tsn}, we conduct reproducible experiments to measure reliability as packet drop rate, latency as \gls{rtt}, and jitter, which are relevant metrics as outlined in the \gls{tsn} methodology \cite{9910175}.
We determine the optimal configuration for a given setup through an iterative investigation of different parameter values.
The experiments are conducted on two \gls{p2p} setups with different hardware capabilities and a bridged setup involving a Linux switch.

Real-time applications have different requirements of reliability, latency, and jitter \cite{9809824}.
Our study encompasses relevant use cases, such as tactile interaction, safety monitoring and control alarms, automated guide vehicles, smart grid protection, and motion control.
Among these, motion control demands the most stringent requirements, such as a maximum latency of \SI{0.1}{\milli\second}, maximum jitter of \SI{0.05}{\milli\second}, and \SI{99.999}{\%} reliability \cite{9809824}.
Our results show that the open-source \gls{opc} \gls{ua} on \gls{cots} hardware meets such requirements.
\section{Background}
\label{sec:background}
This section provides the background of our research, starting with \gls{opc} \gls{ua} \gls{pubsub}.
Related to \gls{tsn}, we focus on \gls{ptp} \cite{9120376}, IEEE 802.1Qav \cite{5375704}, and IEEE 802.1Qbv \cite{8613095} with their corresponding Linux implementations.

\subsection{OPC UA PubSub}
In 2018, the \gls{opc} foundation introduced the \gls{pubsub} extension as part of the \gls{opc} \gls{ua} specification \cite{IEC62541}.
\gls{opc} \gls{ua} \gls{pubsub} encompasses three communication parties.
A publisher sends messages to subscribers through a middleware.
While the \gls{pubsub} protocol does not specify the middleware itself, it relies on underlying protocols for its operation.
Additionally, \gls{opc} \gls{ua} \gls{pubsub} enables encryption and signing of the messages, which are critical for security.

The exchanged messages, \textit{NetworkMessage}, consist of a header and payload organized into \textit{DataSetMessages}.
Each \textit{DataSetMessage} contains a \textit{DataSet}, which is not parsable or included in the message but must be known by the subscriber.
To address this, metadata, including field names and data types, is introduced to provide a clear understanding of the \textit{DataSet}.
This metadata can be obtained through server/client \gls{opc} \gls{ua} or using special mechanisms offered by \gls{opc} \gls{ua} \gls{pubsub}.
The metadata version is an integral part of the \textit{DataSetMessage} to ensure that the publisher and subscriber understand the exchanged messages.

\gls{opc} \gls{ua} \gls{pubsub} messages can be mapped to JSON or binary \gls{uadp}.
While JSON is a standard format for data representation, it is not intended for real-time communication.
Therefore, \gls{opc} \gls{ua} \gls{pubsub} introduces a new application layer protocol named \gls{uadp}, which offers a unique message structure. 
\Cref{fig:uadp-header} depicts the structure of \gls{uadp} without security headers.

\begin{figure}
\includegraphics[width=\columnwidth]{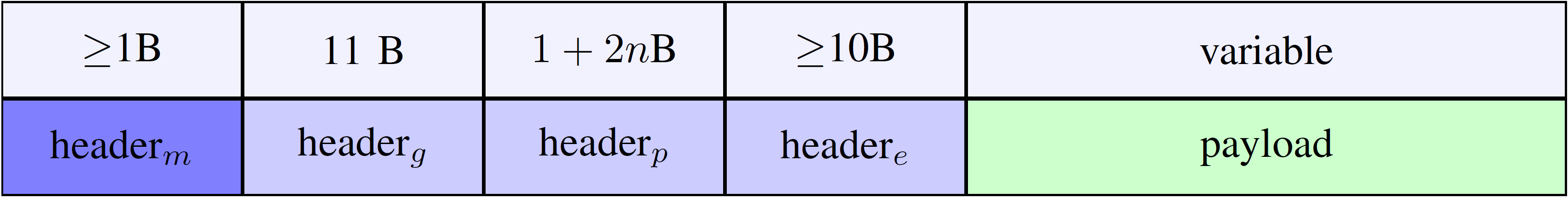}
\caption{UADP message structure \cite{IEC62541}}
\label{fig:uadp-header}
\end{figure}

The message header, denoted as \textit{header$_m$}, is a mandatory component of the communication protocol.
On the other hand, the group header \textit{header$_g$}, payload header \textit{header$_p$}, and extended header \textit{header$_e$} are optional components.
The main header holds information such as the protocol version, flags, the publisher's unique identifier, and the class of \textit{DataSet}.
The group header, in turn, includes the message and sequence numbers, while the extended header includes a timestamp.
The optional payload header encodes the ID of the \textit{DataSetWriter}, allowing subscribers to identify the message's origin and the content of the \textit{DataSet}.
It is important to note that the header is not encrypted, enabling subscribers to filter out unwanted messages.
The structure of encrypted and signed messages is illustrated in \Cref{fig:uadp-header-sec}.

\begin{figure}
\includegraphics[width=\columnwidth]{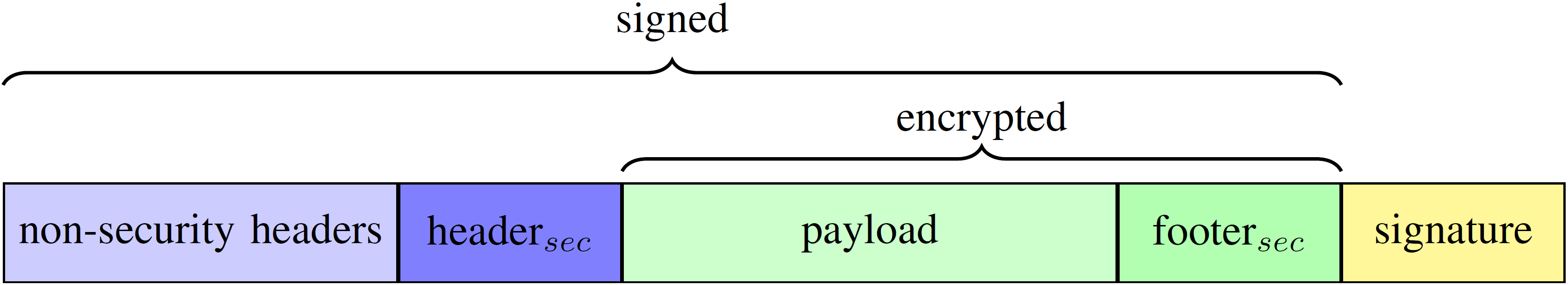}
\caption{UADP security \cite{IEC62541}}
\label{fig:uadp-header-sec}
\end{figure}

\gls{uadp} operates as an application layer protocol.
It can be deployed on top of the following protocols:
\begin{itemize}
    \item Ethernet (layer 2),
    \item \gls{udp} (layer 4),
    \item Advanced Message Queuing Protocol, or \gls{mqtt} (application layer).
\end{itemize}
The implementation of \gls{opc} \gls{ua} \gls{pubsub} varies based on the underlying protocol.
For instance, where \gls{mqtt} is the underlying protocol, \gls{opc} \gls{ua} \gls{pubsub} uses brokers.
Alternatively, when \gls{udp} or Ethernet is used, \gls{opc} \gls{ua} \gls{pubsub} operates brokerless.
Additionally, \gls{opc} \gls{ua} \gls{pubsub} is compatible with \gls{tsn} when deployed with Ethernet.
Due to the support of \gls{tsn}, we focus only on brokerless \gls{opc} \gls{ua} over Ethernet. 

\gls{opc} \gls{ua} over Ethernet is identified by the \texttt{EtherType} \textbf{0xb62c} and encapsulates \gls{uadp} messages directly, bypassing the need for network or transport layer headers.
The addressing format for \gls{opc} \gls{ua} over Ethernet follows the pattern of \texttt{opc.eth://host[:VLAN~ID[.VLAN~priority]]}.
The host component can be a hostname, a \gls{mac} address, or an \gls{ip} address.
When specifying a \gls{mac} address, the bytes should be separated by hyphens instead of colons.
It is important to note that \gls{ip} addresses and hostnames must be resolved to corresponding MAC addresses before communication.

The open62541 project\footnote{\url{https://github.com/open62541/open62541}} is the most widely used open-source implementation of \gls{opc} \gls{ua} \cite{9212091} and the only one that supports the \gls{pubsub} protocol over \gls{tsn} \cite{muhlbauer2021feature}.
It includes an \gls{opc} \gls{ua} stack, a server, a client software development kit, and the implementation of \gls{opc} \gls{ua} \gls{pubsub}.
The application supports JSON or \gls{uadp} encoding over Ethernet or \gls{udp}.

With raw Ethernet as its underlying protocol, \gls{opc} \gls{ua} \gls{pubsub} supports \gls{tsn}.
Since being released before integrating the \gls{taprio} \gls{qdisc} into the Linux kernel, the application implements the IEEE 802.1Qbv standard in software.
The \gls{pubsub} setup requires two hosts: the publisher \textbf{P} and loopback \textbf{L}.
The \gls{opc} \gls{ua} server periodically retrieves and increments variables using an \emph{application thread}.
Additionally, two supplementary threads are responsible for publishing and subscribing.
The L application subscribes to P, storing the subscribed values within its \gls{opc} \gls{ua} server.

A configurable cycle time $c$ defines the duration of a cycle, which corresponds to the \gls{taprio} cycle time.
The publish and subscribe applications begin their cycles at full seconds.
At $0.4 c$ before the start of the next cycle time, the publisher thread is activated and initiates publication.
The transmission time is set to the start of the next cycle time plus a configurable offset $o$. 
\Cref{fig:imp-cycletime} illustrates the starting time of each thread and the offset relative to the cycle time.
The subscriber thread is executed at the start of the cycle ($0.0c$).
At $0.3c$, the user thread stores the values received by the subscriber in the \gls{opc} \gls{ua} address space of the loopback host or increments the variables in the publisher host.
At $0.6c$, $0.4c$ before the start of the next cycle, the publisher thread is again activated, and the execution follows as described above.

\begin{figure}
\includegraphics[width=\columnwidth]{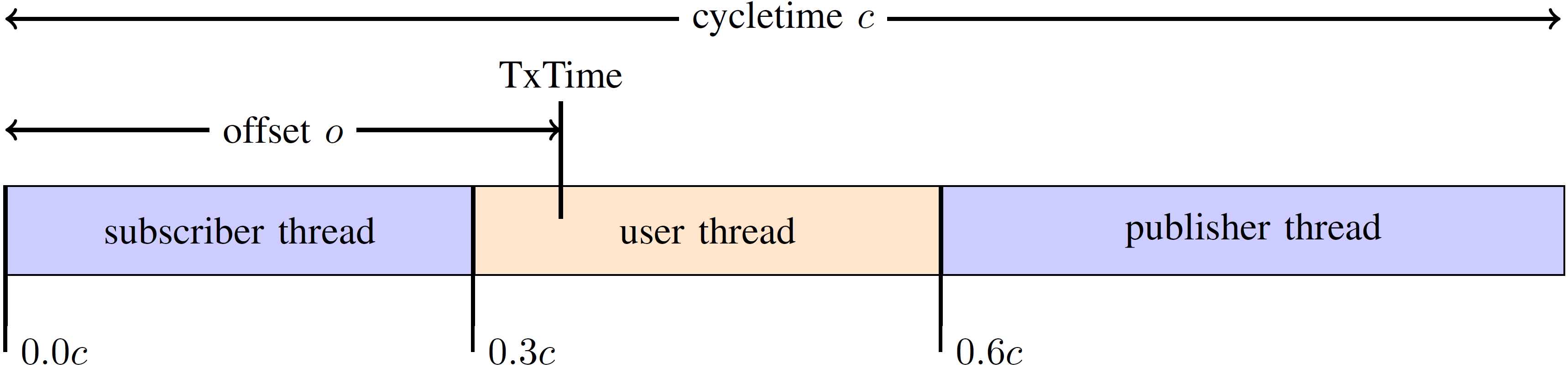}
\caption{Overview of one cycle time}
\label{fig:imp-cycletime}
\end{figure}

\subsection{Time Synchronization}
\gls{ptp} is crucial for synchronous \gls{tsn} standards, such as the IEEE 802.1Qbv. 
It synchronizes individual physical clocks on various hops in high precision.

The Linux \gls{ptp} project\footnote{\url{https://sourceforge.net/projects/linuxptp/}} encompasses the tools \textit{ptp4l}\footnote{\url{https://linux.die.net/man/8/ptp4l}}, \textit{phc2sys}\footnote{\url{https://linux.die.net/man/8/phc2sys}}, and \textit{pmc}\footnote{\url{https://linux.die.net/man/8/pmc}}.
The \textit{ptp4l} tool, a command-line utility, implements the \gls{ptp} standard IEEE 1588 \cite{9120376} and can operate over Ethernet, IPv4, or IPv6 networks.
The \textit{ptp4l} daemon must run on all interfaces to synchronize the system's clocks and identify the \gls{ptp} \gls{gm}, which serves as the reference time for the entire system organized in a master-slave hierarchy.
The \textit{phc2sys} tool synchronizes the clocks within a single system and can run in automatic mode, utilizing information from the \textit{ptp4l} daemon to achieve synchronization.
Hardware timestamping can be leveraged to achieve nanosecond-level precision when utilizing a \gls{nic} that supports IEEE 802.1AS, such as the commercially available Intel\textsuperscript{\textregistered} I210 \gls{nic}.

\subsection{Linux Traffic Control}
We provide an overview of the Linux \gls{tsn} implementations based on the details described in the IEEE standards. 
Within the scope of this paper, we focus on the \gls{mqprio} and \gls{taprio} \glspl{qdisc} as parents and the \gls{etf} and \gls{cbs} \glspl{qdisc} as children, specifically in the context of \gls{tsn}.

The IEEE 802.1Qav standard outlines two algorithms for shaping and prioritizing network traffic.
The first algorithm, strict priority forwarding, prioritizes traffic based on class priority, with higher priority classes transmitted first.
This algorithm is already specified in the IEEE 802.1Q standard \cite{6009146}.
As part of IEEE 802.1Qav, it is considered a \gls{tsn} algorithm.
If no packets are in the higher priority class, lower priority packets are transmitted.
This algorithm is implemented with the \gls{mqprio} \gls{qdisc}\footnote{\url{https://man7.org/linux/man-pages/man8/tc-mqprio.8.html}}.
\glspl{qdisc} handles traffic and is organized in a parent-child hierarchy.
Relevant parameters are \texttt{num\_tc}, which indicates the number of configured traffic classes, and \texttt{map}, which maps packet priorities to corresponding traffic classes.
The second algorithm defined by IEEE 802.1Qav is the \gls{cbs} algorithm.
Bridges implementing IEEE 802.1Qav should prioritize \gls{cbs} traffic classes over those using strict priority forwarding, as indicated by the default \gls{pcp} to \gls{tc} mapping in \Cref{tab:pcp-tc}.

\begin{table}
\centering
\caption{PCP to TC mapping according to IEEE 802.1Qav}
\label{tab:pcp-tc}
\begin{tabular}{l c c c c c c c c}
\textbf{PCP} & 0 & 1 & 2 (SR-B) & 3 (SR-A) & 4 & 5 & 6 & 7 \\ \midrule
\textbf{TC} & 1 & 0 & 6 & 7 & 2 & 3 & 4 & 5 \\
\end{tabular}
\end{table}

The \gls{cbs} \gls{qdisc}\footnote{\url{https://man7.org/linux/man-pages/man8/tc-cbs.8.html}} regulates and secures bandwidth allocation for a specific traffic class.
It is implemented as a child \gls{qdisc} in conjunction with a root \gls{qdisc}, such as \gls{mqprio}, which performs traffic classification.
\Cref{fig:cbs} shows the credit lifecycle with the four \gls{cbs} parameters - $highCredit_x$, $lowCredit_x$, $idleSlope_x$, and $sendSlope_x$. 
The first two define the maximum and minimum allowed credit.
$idleSlope_x$ marks the credit replenishment rate, and $sendSlope_x$ the credit spending rate.
Once a frame arrives, the credit starts to build up \raisebox{.5pt}{\textcircled{\raisebox{-.9pt} {1}}}. 
After the $highCredit_x$ is reached \raisebox{.5pt}{\textcircled{\raisebox{-.9pt} {2}}}, the frame is sent after non-policed frames are transmitted, following the $sendSlope_x$ parameter.
If no further frames are in the queue, the credit drops to 0, \raisebox{.5pt}{\textcircled{\raisebox{-.9pt} {3}}}. 
If there is still some credit from $sendSlope_x$, the next frame gets sent out immediately. 
This triggers the credit rebuilt since the credit is fully depleted, \raisebox{.5pt}{\textcircled{\raisebox{-.9pt} {4}}}. 
It is worth mentioning that the parameters in \gls{cbs} are defined in bits or bits per second, as specified by the IEEE 802.1Qav standard, as opposed to bytes or kilobits per second in Linux.

\begin{figure}
\includegraphics[width=\columnwidth]{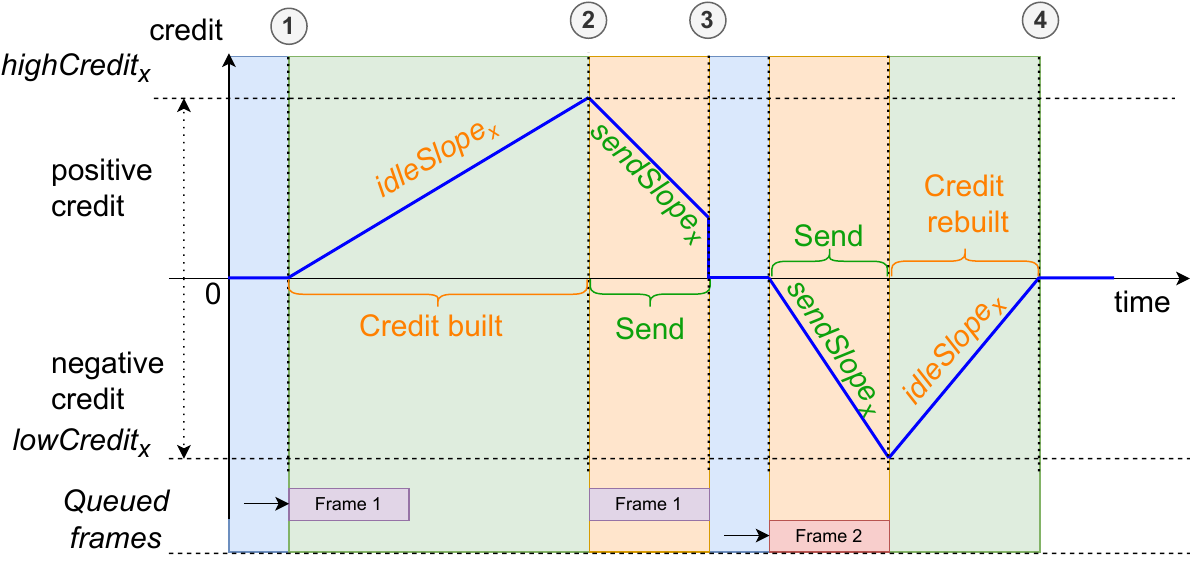}
\caption{Credit of \gls{cbs} over time, based on\cite{9910175,5375704}}
\label{fig:cbs}
\end{figure}

In 2015, IEEE published the IEEE 802.1Qbv standard, which provides enhancements for scheduled traffic.
With the implementation of IEEE 802.1Qbv, it is possible to implement a Time-Division Multiple Access scheme for Ethernet networks.
Each traffic class has a queue with an associated \emph{gate}, which can be opened or closed anytime.
Packets can only be dequeued from a queue when its gate is open.
The gates are controlled by a schedule consisting of gate operation instructions.
Each operation specifies a gate configuration and the duration for which the configuration remains valid.
The operations are executed in sequence, taking into account the specified intervals.
This process is repeated to form a cycle, with the cycle time equal to the sum of all the intervals in the schedule.

The \gls{taprio} \gls{qdisc}\footnote{\url{https://man7.org/linux/man-pages/man8/tc-taprio.8.html}}, also known as the \gls{tas}, is designed to calculate schedules with time slots for different traffic classes as specified in the IEEE 802.1Qbv standard.
Like the \gls{mqprio} \gls{qdisc}, packets are initially mapped to traffic classes and Tx queues.
To synchronize schedules across all network devices, it is necessary to correctly set the \texttt{base-time} and ensure its alignment among all hosts.
As a result, \gls{taprio} falls under synchronous \gls{tsn} standards.
The base-time is specified in nanoseconds and indicates the starting point of the schedule.
The \texttt{sched-entry} parameters follow and define the gate state and its duration.
The syntax is  \texttt{sched-entry S \$MASK \$DURATION}, where \texttt{\$DURATION} represents the time window duration, and \texttt{\$MASK} indicates which gate is open or closed during that window.
To avoid interference between windows, guard windows can be added between them, with the guard window size computed based on the packet serialization time, packet size, and link speed.
The cycle time is the sum of all schedule entries, including the guard windows.
The \texttt{flags} support TxTime mode (\texttt{flags 0x1}) and full offload mode (\texttt{flags 0x2}).
The TxTime mode sets the packet's TxTime and uses the \gls{etf} \gls{qdisc} to control when a packet is transmitted.
The \texttt{txtime-delay} value is required when the flags are set to \texttt{0x1} and serves as a delay to compensate for system delay.
It should be set to a value greater than the $\delta$ value of the \gls{etf} \gls{qdisc}.

The \gls{etf} \gls{qdisc} is employed when a packet is transmitted at a specific TxTime\footnote{\url{https://man7.org/linux/man-pages/man8/tc-etf.8.html}}.
This feature, the \emph{LaunchTime}, is illustrated in \Cref{fig:etf}.
In \Cref{fig:etf}a, a packet with priority three and TxTime \textit{T} arrives at the root \gls{qdisc} for further processing.
In \Cref{fig:etf}b, the packet is classified based on its priority to a given class and corresponding child \gls{qdisc} queue.
The packet is dequeued at time $delta$ before the TxTime to the ring buffer, where it is taken to the \gls{nic}, \Cref{fig:etf}c.
At the TxTime, the packet is eventually dequeued to the wire, as shown in \Cref{fig:etf}d.

\begin{figure*}
\includegraphics[width=\textwidth]{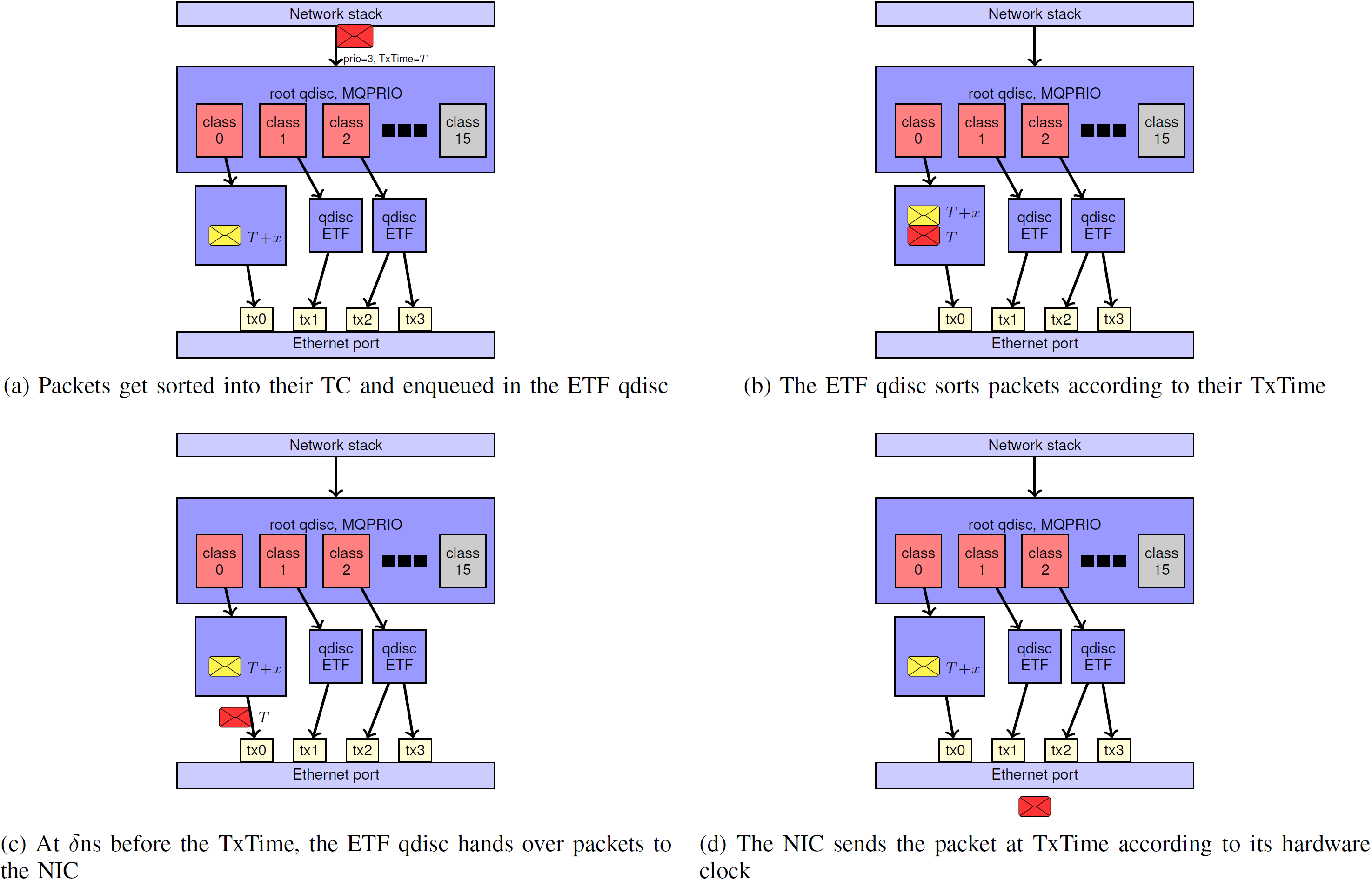}
\caption{Linux Traffic Control at the example of \gls{mqprio} and \gls{etf}}
\label{fig:etf}
\end{figure*}

As shown in the example, \gls{etf} can be utilized as a child \gls{qdisc} with \gls{mqprio} or mainly with the \gls{taprio} as the parent \gls{qdisc}.
It is especially crucial for the \gls{taprio} \gls{qdisc}, as the packets could only be appropriately dequeued within their respective windows with it.
The TxTime for a packet is specified in the SKB SO\_TXTIME option.
If packets arrive at the queue after their designated TxTime, they are dropped by the \gls{nic}.
The \gls{etf} \gls{qdisc} sorts packets based on their TxTime, and at a configurable offset $\delta$ before the TxTime, the \gls{qdisc} dequeues the packet for transmission to the \gls{nic}.
The offset $\delta$ can also compensate for system delay.
If the \gls{nic} supports the \emph{LaunchTime} feature, the \gls{etf} can be offloaded for improved performance.

Linux has the \gls{tc} command line tool to manage the configuration of the \glspl{qdisc} in the networking stack\footnote{\url{https://linux.die.net/man/8/tc}}.
The traffic packet priorities are mapped to traffic classes corresponding to one or more Tx queues.
The Tx queue with the lowest number is emptied first.
The Intel\textsuperscript{\textregistered} I210 \gls{nic}, for example, has four hardware queues, each of which can have its child \gls{qdisc} configured.
The first \gls{qdisc} corresponds to hardware queue one and is assigned the highest traffic priority.
Priorities are stored in the Linux \gls{skb} and correspond to their VLAN \gls{pcp} header field, as described in \cite{8502479}.
If a \gls{nic} supports the respective standards, the algorithms can be offloaded, potentially resulting in a processing speedup.

\section{Related Work}
\label{sec:related}
Two similar related works \cite{8502479} and \cite{9212171} employ \gls{opc} \gls{ua} \gls{pubsub} to establish communication between two \gls{p2p} connected devices, utilizing the Intel\textsuperscript{\textregistered} I210 \gls{nic} and the open-source \textit{open62541} stack.
In \cite{8502479}, the \gls{taprio} cycle time is set to \SI{100}{\microsec}, while in \cite{9212171}, it is set to \SI{200}{\microsec}.
Our experiments use a cycle time of \SI{250}{\microsec}, which is determined by the system performance.
\Cref{sc:results-paramsandmetrics} outlines determining the optimum cycle time value and other related parameters.

Research by Eckhardt et al. \cite{8869060} utilizes specialized embedded hardware for \gls{tsn} instead of \gls{cots} hardware.
The \gls{opc} \gls{ua} software stack is not specified.
\gls{tas} is preferred without providing any information regarding the implementation.
In contrast to our study, \cite{8869060} is configured with shorter cycle times for \gls{tas} than for the application.
Moreover, it uses software timestamps, which reduce the accuracy of measurements but allow the investigation of application bottlenecks.

A study by Farzaneh et al. \cite{8275648} conducts \gls{tas} experiments using cyclic real-time traffic with a cycle time of \SI{500}{\microsec}.
The setup employs commercial switches with a \gls{tas} implementation in a field programmable gate array and uses three sequentially connected switches.
However, since the end devices lack hardware timestamping, the timestamps of the adjacent switch ports are utilized.
The traffic is unidirectional, with the first switch acting as the source and the second as the sink.
In contrast, our study uses a loopback application on the second host and mirrors the messages to their origin, allowing for the \gls{rtt} analysis.
In our research, we also employ \gls{mqprio}, the Linux implementation of strict priority forwarding, to perform experiments with priority forwarding, similar to the approach taken by \cite{8275648}.

Li et al. \cite{9247090} investigate a bridged \gls{tas} network utilizing two specialized \gls{tsn} switches and hardware \gls{tsn} modules to enable \gls{tsn} on end devices.
However, the setup relies solely on commercial hardware, contrasting our approach.
A cycle time of \SI{20}{\milli\second} is used, and all \gls{tsn} traffic is transmitted in a single direction.
The implementation has \gls{opc} \gls{ua} with a server/client communication pattern instead of \gls{pubsub}.
Only the client is open-source.
The study does not include the jitter or packet drop rate measurements.
Although the implementation achieves bounded latency using \gls{tas}, the bound is almost \SI{900}{\microsec}, significantly higher than the average latency of approximately \SI{40}{\microsec} without \gls{tas}.

A study by Gogolev et al. \cite{8502597} investigates the performance of \gls{tas} switches based on proprietary software using specialized evaluation hardware.
The approach utilizes the \gls{opc} \gls{ua} server and client to perform read or write requests at varying intervals.
However, unlike our \gls{tas} configuration, the work does not examine worst-case \gls{rtt}, jitter, or packet drop, focusing solely on average \gls{rtt}.
A subsequent study by the same authors \cite{8972252} combines \gls{tas} with \gls{cbs} to limit bandwidth for \gls{be} traffic.
The findings reveal that the impact of \gls{tas} is more significant than that of \gls{cbs}.

Arestova et al. \cite{arestova2021service} research focuses on the gls{taprio} framework and experiments on a two-node network.
The experiments analyze the performance of a one-way data flow from a sender to a receiver.
The sender uses a specialized \gls{tsn} \gls{nic} from Kontron, and the receiver uses an Intel\textsuperscript{\textregistered} I210 \gls{nic} to capture hardware timestamps.
Our work shares the same \gls{opc} \gls{ua} \gls{pubsub} implementation as \cite{arestova2021service}, \textit{open62541}, and uses a \SI{1}{\milli\second} cycle time with a \SI{100}{\microsec} priority traffic window.

An evaluation by Gruener et al. \cite{9921503} uses \gls{opc} \gls{ua} \gls{pubsub} over \gls{tsn} on \gls{cots} hardware.
Similarly, like \cite{8502479}, it relies on the \textit{open62541} stack.
Concerning network size, the evaluation only considers measurements in \gls{p2p} topology using Intel\textsuperscript{\textregistered} I350 \gls{nic} that supports \gls{ptp}.
The setup uses the real-time kernel on one of the hosts to limit operating system noise.
Since the \gls{nic} does not support any additional \gls{tsn} features, the focus is restricted to the \gls{taprio} offered by the \gls{opc} \gls{ua} application.
The implementation deploys Xpress Data Path to improve the loopback performance for faster packet processing.

Finally, Denzler et al. \cite{9779177} focus on the \textit{open62541} \gls{opc} \gls{ua} \gls{pubsub} stack extension with the 802.1q VLAN tag.
It facilitates IEEE 802.1Qbv time-aware scheduling.
This work investigates end-to-end timing measures and worst-case execution time analyses, considering various payloads.
The experimental setup is distinguished using time-predictable T-CREST platforms hosting the publisher and subscriber with a TSN network handling message transmission.
Our work diverges from \cite{9779177} as we implement our timing analyses using \gls{cots} hardware.
Furthermore, we expand our scope to cover all Linux TSN queuing disciplines rather than exclusively focusing on IEEE 802.1Qbv time-aware scheduling, thus adding another dimension to our comparative analysis.

The position of this work and the research gap in the existing literature are presented in \Cref{tab:related}.
We aim to analyze all \glspl{qdisc} using the same open-source \gls{opc} \gls{ua} software and \gls{cots} hardware.
This standardized approach enables a fair comparison of \glspl{qdisc} by subjecting them to identical testing conditions.
 
\begin{table}
\centering
\caption{Position of our work}
\label{tab:related}
\resizebox{\columnwidth}{!}{
\begin{tabular}{l c c c c c c c c c}
Works  & \makecell{\cite{8502479} \\ \cite{9212171}} & \cite{8869060} & \cite{8275648} & \cite{9247090}  & \makecell{\cite{8502597} \\ \cite{8972252}} & \cite{arestova2021service} & \cite{9921503} & \cite{9779177} & \makecell{this \\ work} \\ \toprule

\gls{opc} \gls{ua} p/s           & \checkmark& \checkmark & - & - & - & \checkmark & \checkmark & \checkmark & \checkmark \\ 
P2P                 & \checkmark  & \checkmark & - & - & - & \checkmark & \checkmark & \checkmark & \checkmark \\ 
Bridge              & -           & -          & \checkmark& \checkmark & \checkmark & -  & - & \checkmark & \checkmark  \\ 
ETF                 & \checkmark  & -          & -         & -          & -          & -  & - & - & \checkmark \\ 
CBS                & -           & -          & -         & -          & \checkmark & -  & - & - & \checkmark \\ 
TAPRIO              & -           &(\checkmark)& \checkmark& \checkmark & \checkmark & \checkmark & \checkmark & \checkmark & \checkmark\\ 
COTS HW             &\checkmark   & -          & -         & -          & -          & -          & \checkmark & - & \checkmark\\ 
Open-source         & \checkmark  & ?          & -         & client     & -          & \checkmark & \checkmark & \checkmark & \checkmark\\ \bottomrule
\end{tabular}
}
\end{table}
\section{Experimental Setting}
\label{sec:experimental_setting}

This section describes the setting of the experiments.
We outline the measurement setup, including the hardware and software components.
This section provides a detailed explanation, ensuring experiments' repeatability and reliability.

\subsection{Hardware \& Software Setup}

Our evaluation consists of three setups designed to test different aspects, as illustrated in \Cref{fig:testSetups}.
\Cref{fig:setup-e3} and \Cref{fig:setup-d} represent \gls{p2p} topologies, with the difference being the inclusion of an additional dedicated link for clock synchronization between the nodes in \Cref{fig:setup-d}.
This dedicated link is required due to an existing problem of \gls{taprio} with \gls{etf} and \gls{ptp} \cite{10150312}.
In contrast, \Cref{fig:setup-b} depicts the bridged topology.

\begin{figure}
\centering
\subfloat[Setup E3\label{fig:setup-e3}]{\includegraphics[trim={0 2.8cm 48cm 0},clip,width=0.32\columnwidth]{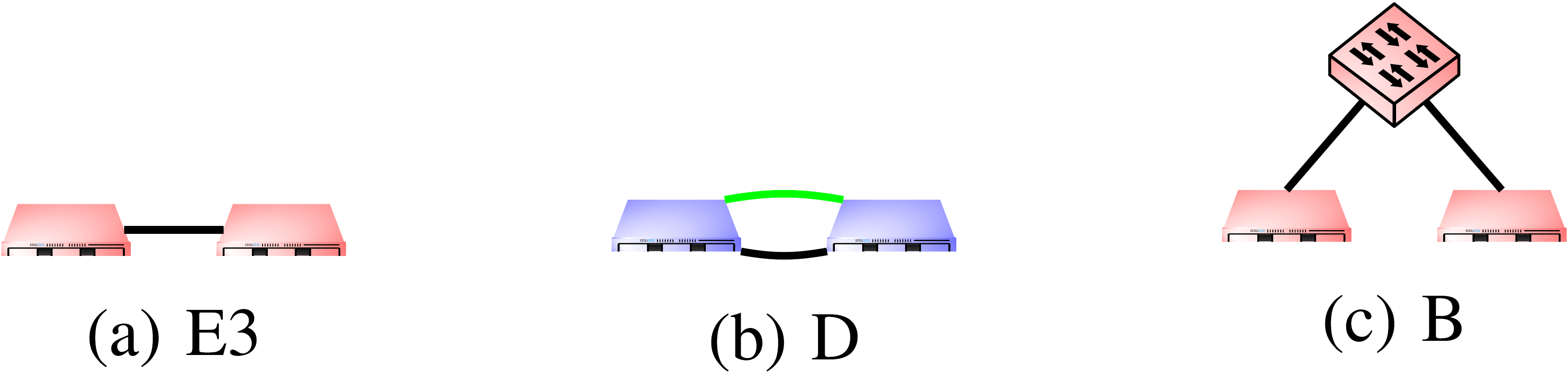}}
\subfloat[Setup D\label{fig:setup-d}]{\includegraphics[trim={24cm 2.8cm 24cm 0},clip,width=0.32\columnwidth]{figures/kirda6.png}}
\subfloat[Setup B\label{fig:setup-b}]{\includegraphics[trim={48cm 3.2cm  0 0},clip,width=0.32\columnwidth]{figures/kirda6.png}}
\caption{Test setups with standard hosts (red), powerful hosts (blue), and a dedicated \gls{ptp}-link (green)}
\label{fig:testSetups}
\end{figure}

\Cref{tab:hpclpc} shows the hardware specifications for each setup. 
Setups E3 and B rely on the Intel\textsuperscript{\textregistered} Xeon\textsuperscript{\textregistered} E3-1265L V2 CPU and \SI{16}{\giga\byte} \gls{ram}.
On the other hand, setup D uses Intel\textsuperscript{\textregistered} Xeon\textsuperscript{\textregistered} D-1518 CPU with \SI{128}{\giga\byte} \gls{ram}.
All setups are interconnected with Intel\textsuperscript{\textregistered} I210 \gls{nic} that supports 1GbE Ethernet and complies with IEEE 802.1AS, IEEE 802.1Qav, and IEEE 802.1Qbv.
All selected components are easily accessible and belong to the \gls{cots} hardware segment.

\begin{table}
\centering
\caption{Hardware specification of setups E3, B and D}
\resizebox{\columnwidth}{!}{
\label{tab:hpclpc}
\begin{tabular}{lcc}
\toprule
 & Setup E3 and B & Setup D\\ \midrule
CPU     & 4C Intel\textsuperscript{\textregistered} Xeon\textsuperscript{\textregistered} E3-1265L V2 & 4C Intel\textsuperscript{\textregistered} Xeon\textsuperscript{\textregistered} D-1518 \\                   
\gls{ram}     & 16\,GB DDR3 & 128\,GB DDR4  \\
NIC    & 4\,$\times$\,1 GbE Intel\textsuperscript{\textregistered} I210$^\dagger$  & 6\,$\times$\,1 GbE Intel\textsuperscript{\textregistered} I210$^\dagger$ \\
Motherboard & ASRock Z77E-ITX & Supermicro X10SDV-TP8F\\
\bottomrule
\multicolumn{3}{l}{$^\dagger$IEEE 802.1Qav, Qbv, AS}
\end{tabular}
}
\end{table}

To ensure experiment repeatability, nodes use live images that run exclusively in \gls{ram}, ensuring that all residual states are erased after each reboot.
All hosts run the Linux kernel 5.4.0-45 with an RT-PREEMPT patch, allowing us to use CPU isolation to prevent the system scheduler from placing other tasks on the defined CPU core.
To achieve this, we manually set the number of cores not interrupted by other threads.
We use the plain orchestration service \cite{gallenmuller2021pos} to orchestrate all experiments.

In the bridged topology, we capture timestamps using \textit{tcpdump}\footnote{\url{https://www.tcpdump.org/}}, which allows for high-precision hardware timestamping if the \gls{nic} supports it \cite{rezabek2022engine}.
\Cref{fig:measurementSetup} shows the measurement points of the software timestamp with the black dots and the hardware timestamp with the red diamonds.
We also collect time measurements in the application developed by \cite{8502479}, represented by green squares.
However, we find that these timestamps have lower accuracy and thus are not further considered.
\textit{tcpdump} records timestamps on the bridge only in the ingress direction.
Consequently, we do not include measurements of egress traffic on the bridge.
Moreover, we exclude non-solid edges in \Cref{fig:measurementSetup} for the \gls{rtt} measurement, as they rely on software timestamping, which results in lower precision.

\begin{figure}
\includegraphics[width=\columnwidth]{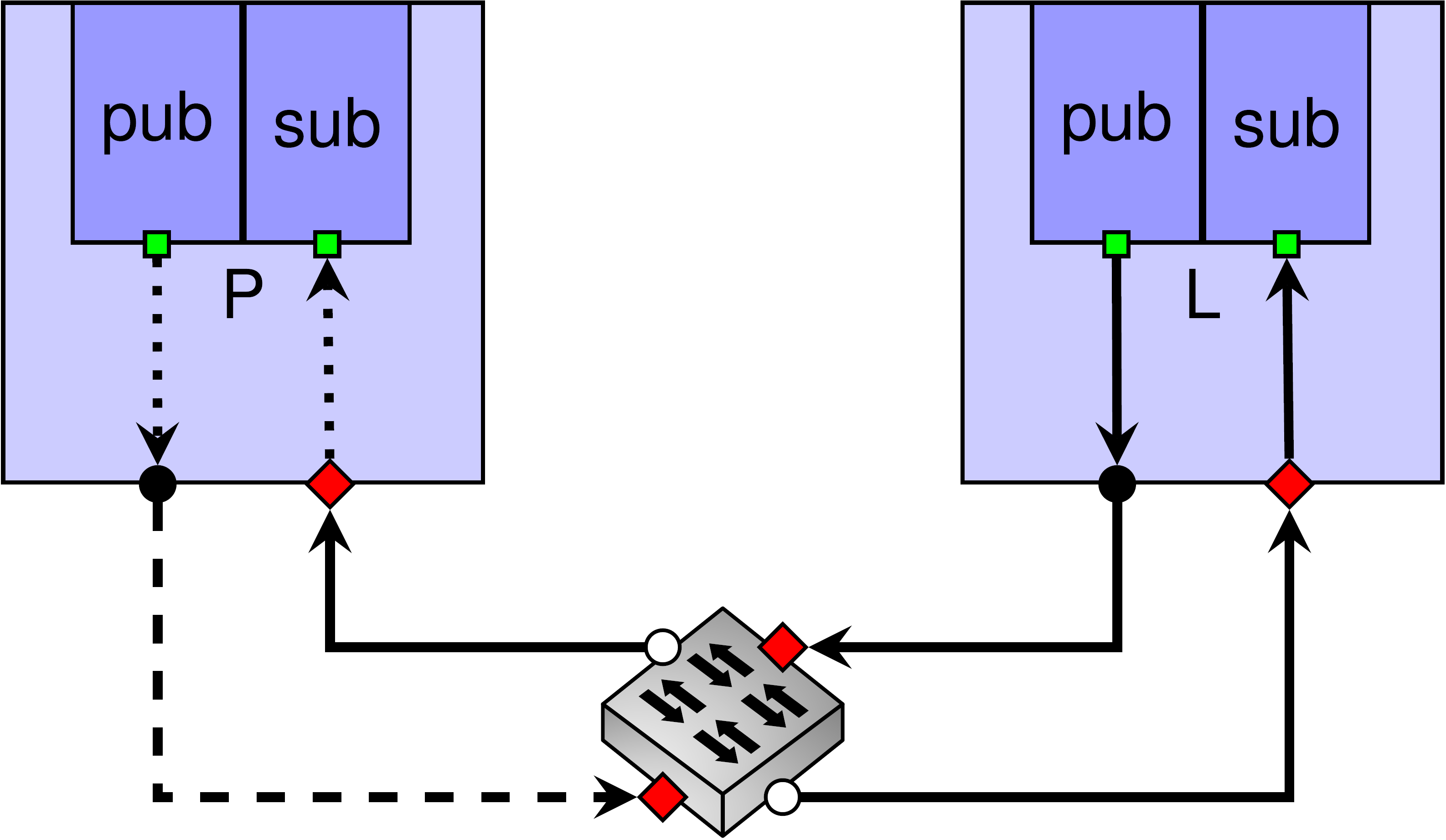}
\caption{Measurement points of the bridged topology}
\label{fig:measurementSetup}
\end{figure}

\subsection{PTP Configuration}

Our setups rely on the \textit{ptp4l} for time synchronization.
This daemon synchronizes the hardware clock of the interface to a \gls{gm} clock, which ensures accurate and consistent timekeeping across all devices in the network.
Once \textit{ptp4l} has synchronized the hardware clock, we use the \textit{phc2sys} tool to synchronize the system clock to the hardware clock.
This ensures that our system clock is also accurately synchronized to the \gls{gm} clock, essential for reliable data transfer and analysis.

In the \gls{p2p} topology, we configure the publisher host as the \gls{gm} and the loopback host as a slave.
This means that the publisher host is responsible for providing the reference time to the loopback host, which ensures that both hosts are synchronized to the same clock.
In the bridged topology, we configure the bridge as the \gls{gm} and the publisher and loopback hosts as slaves.
This ensures that the distance to the next \gls{gm} clock is a maximum of one hop, which helps to minimize latency and maintain accurate timekeeping across all devices in the network \cite{9964658}.

\subsection{Configuration of qdiscs}

The most straightforward \gls{tsn} configuration at the bridge is \gls{mqprio}.
It allows for the mapping of desired priorities to class mapping.
To maintain consistency throughout the tests, we adopt a uniform mapping where the highest priority is assigned to \gls{opc} \gls{ua} traffic and is mapped to the first hardware queue.
All other priorities are treated as \gls{be} traffic and are forwarded with the lowest priority.

We use the \gls{etf} \gls{qdisc} in offload mode to leverage the LaunchTime feature of the Intel\textsuperscript{\textregistered} I210 \gls{nic}.
We configure it as the child \gls{qdisc} of the parent \gls{mqprio} \gls{qdisc}.
This configuration lets us prioritize traffic and effectively manage network resources during data transfer.
To ensure the proper functioning of the \gls{etf}, we install it following the guidelines provided by the manufacturer.

\gls{cbs}, similar to \gls{etf}, can only be used as a child \gls{qdisc} and applied to the first two hardware queues on the Intel\textsuperscript{\textregistered} I210 \gls{nic}.
Four specific parameters are required to configure \gls{cbs}: $idleSlope_x$, $sendSlope_x$, $highCredit_x$, and $lowCredit_x$.
These parameters can be set according to the guidelines outlined in the IEEE 802.1Qav standard, as discussed in \Cref{sec:background}.

We utilize \gls{taprio} to classify packets into different traffic classes.
To ensure consistency across all hosts, we use a base-time of 1 second and share the same schedule entries.
Since the \gls{opc} \gls{ua} application initiates its first cycle at an integer number of seconds, \gls{taprio} 's and its cycles also start simultaneously.
Our schedule begins with a \gls{be} traffic window, the offset-window, since its duration corresponds to the configurable offset in the \gls{opc} \gls{ua} application.
In TxTime-assisted mode, \gls{taprio} configures a TxTime for each packet using the 0x1 flag.
It is accounted for the maximum delay between \gls{taprio} and the \gls{nic}.
Additionally, we use \gls{etf}-assisted mode, which involves sending packets at precisely the TxTime if a child \gls{etf} \gls{qdisc} with activated offloading is installed for a traffic class.
The child \gls{etf} \gls{qdisc} must be configured with skip sock check.
Otherwise, \gls{etf} would drop packets from \gls{taprio}.
\section{Results}
\label{sec:results}
This section presents the results obtained from the experiments.
Firstly, we determine the scheduling latency of our hosts, which is the duration between the requested wake-up time of a thread and the actual scheduling by the operating system.
Subsequently, we conduct experiments on three different setups: E3 and D, consisting of two hosts connected \gls{p2p} and B, including a bridge with additional \gls{be} traffic.
For the \gls{p2p} setups, we evaluated all available \gls{tsn} \glspl{qdisc}, namely \gls{mqprio}, \gls{cbs}, \gls{etf}, and \gls{taprio}, alongside the default \gls{qdisc} of Linux, \gls{fqcodel}, as a benchmark.
We conclude by outlining the potential threats to validity.

\subsection{Experiment Parameters}
\label{sc:results-paramsandmetrics}

For each test, we send 700,000 packets and repeat it five times.
We evaluate the packet drop rates, \gls{rtt}, and jitter.
We differentiate between $d_P$ and $d_L$ for packet drop rates, representing rates on the publisher and loopback hosts, respectively, and their sum $d_\Sigma$.
Additionally, for the bridge setup, we represent the drop rates at the bridge to loopback with $d_{b\rightarrow L}$ and to the publisher with $d_{b\rightarrow P}$.

We set up the RT-PREEMPT kernel on all experiment hosts, which results in scheduling latencies between \SI{40}{} to \SI{130}{\microsec}, depending on the setup.
To measure the scheduling latency, we run \textit{cyclictest}\footnote{\url{https://man.archlinux.org/man/cyclictest.8}} on the experiment hosts with the publisher thread's priority and the lowest cycle time used in our experiments as the interval.
We experiment for \SI{60}{\second}, resulting in the values above.
The RT-PREEMPT Linux kernel improves maximum latencies, as evidenced by \SI{7.5}{\milli\second} and \SI{0.24}{\milli\second} latencies for setup E3 and D, respectively, when using a regular kernel.
We use more conservative values to mitigate the risk of dropping a packet before dequeuing.
\Cref{tab:perf-cycles} shows the impact of cycle time on publisher and loopback hosts with and without \gls{cpu} isolation on drop rates, \gls{rtt}, and jitter.
According to the results, \SI{200}{} and \SI{250}{\microsec} values perform the best, considering $d_\Sigma$ and $d_\Sigma$ (isol).
Our parameter study reveals that selecting a lower cycle time resulted in higher drop rates caused by missing the window opening.
In contrast, increasing the cycle time decreases the drop rate and jitter while increasing delay since packets arrive when the window is open and thus do not have to wait for the next window cycle.
More conservative \SI{200}{} and \SI{250}{\microsec} values result in higher delays but lower jitter.
We select a cycle time of \SI{250}{\microsec} since it offers a lower jitter.

\begin{table}
\centering
\caption{Mean results of ETF with increasing cycletimes}
\label{tab:perf-cycles}
\begin{tabular}{l c c c c c}
cycletime [µs]                 &  100 & 125  &  150  & 200   & 250\\ \toprule
$d_P$ [\%]               & 0.029    & 0.185    & 0.363    & 0.037    & 0.162 \\
$d_P$ (isol) [\%]        & 0.0313    & 0.0559    & 0.0092    & 0.0092    & 0.0025 \\
$d_L$ [\%]               & 4.93    & 3.04    & 0.122    & 0.126    & 0.117 \\
$d_L$ (isol) [\%]        & 4.09    & 2.45    & 0.0497    & 0.0481    & 0.154 \\
$d_\Sigma$ [\%]          & 4.96    & 3.22    & 0.485    & 0.163    & 0.279 \\
$d_\Sigma$ (isol) [\%]   & 4.12    & 2.51    & 0.0589    & 0.0573    & 0.156 \\ \midrule
RTT [µs]             & 195 & 129 & 170 & 246 & 266 \\
jitter [µs]         & 10& 1.5 & 0.33 & 0.528 & 0.237 \\ \bottomrule
\end{tabular}
\end{table}

\Cref{tab:perf-offset} shows the drop rate results for offset values ranging from \SI{0}{} to \SI{250}{\microsec}.
The drop rate begins at \SI{0.0772}{\%} for zero offset and peaks at \SI{50}{\microsec}.
Then, the drop rate decreases until it reaches its minimum of \SI{0.0075}{\%} at the offset value of \SI{150}{\microsec}.
For higher offsets, the drop rate increases again. 
The mean jitter follows the same trend, with outliers exhibiting multiples of the cycle time $c$ = \SI{250}{\microsec}.
These outliers are linked to the subscriber thread missing the timeslot, causing the subsequent packet drop.
This behavior is unrelated to sending time precision.
Conversely, a \SI{50}{\microsec} offset demonstrates the opposite behavior.
To ensure consistent behavior, we set the offset value to guarantee that all packets arrive after the subscriber thread reads new messages.
This choice increases latency but reduces the jitter and drop rate.
After analyzing the results, we determine that the offset value of \SI{150}{\microsec} produces the lowest drop rate, average jitter, and the fewest number of jitter and \gls{rtt} outliers.
Therefore, we fix the offset value to \SI{150}{\microsec} for all subsequent experiments.

\begin{table}
\centering
\caption{Mean results of ETF with increasing offset}
\label{tab:perf-offset}
\begin{tabular}{l c c c c c c}
offset [µs]            &  0    & 50    &  100  & 150   & 200   & 250 \\ \toprule
$d_P$ [\%]      & 0.0027    & 0.0029    & 0.0058    & 0.0013    & 0.0039    & 0.0078    \\
$d_L$ [\%]      & 0.0745    & 0.186    & 0.0327    & 0.0061    & 0.0191    & 0.0722    \\
$d_\Sigma$ [\%] & 0.0772    & 0.187    & 0.0385    & 0.0075    & 0.023    & 0.08    \\ \midrule
RTT [µs]    & 250 & 500 & 500 & 500 & 500 & 500 \\
jitter [µs] & 0.471 & 0.844 & 0.274 & 0.216 & 0.265 & 0.449 \\ \bottomrule
\end{tabular}
\end{table}

The next parameter we optimize is the $\delta$ value, which must fall within certain upper and lower bounds.
Specifically, due to the scheduling latency, we set the lower bound at $\delta > \SI{130}{\microsec}$ and the upper bound at the publishing latency $\delta < 0.4c+o = \SI{250}{\microsec}$.
Our findings indicate that the $\delta$ value greater than the publishing latency $l_{pub}$ does not significantly enhance performance.
To ensure regular intervals, we conduct measurements in the range of [125, 250] with steps of \SI{25}{\microsec}.
\Cref{tab:perf-delta} shows the results with increasing $\delta$ values.
The drop rate begins at \SI{0.0203}{\%} for $\delta = \SI{125}{\microsec}$ and peaks at $\delta = \SI{150}{\microsec}$.
The drop rate remains constant for higher $\delta$ values at around \SI{0.07}{\%}.
The mean jitter exhibits a similar pattern to the drop rate.
Notably, the delta value of \SI{200}{\microsec} produces the lowest drop rate, \gls{rtt}, and jitter values.
The \gls{rtt} remains almost constant at \SI{500}{\microsec} and is negligibly affected by $\delta$.
Consequently, we set the $\delta$ value to \SI{200}{\microsec} for all subsequent measurements.

\begin{table}
\centering
\caption{Mean results of ETF with increasing $\delta$}
\label{tab:perf-delta}
\begin{tabular}{l c c c c c c}
$\delta$ [µs]             & 125   & 150   &  175  & 200   & 225   & 250\\ \toprule
$d_P$ [\%]       & 0.0028    & 0.0009    & 0.0012    & 0.0023    & 0.0002    & 0.0005    \\
$d_L$ [\%]       & 0.0175    & 0.15    & 0.0638    & 0.0077    & 0.0675    & 0.0661    \\
$d_\Sigma$ [\%]  & 0.0203    & 0.15    & 0.065    & 0.01    & 0.0677    & 0.0666    \\ \midrule
RTT [µs]     & 500 & 500 & 500 & 500 & 500 & 500 \\
jitter [µs]  & 0.245 & 0.7 & 0.414 & 0.221 & 0.429 & 0.424 \\ \bottomrule
\end{tabular}
\end{table}

\Cref{tab:opt-params} displays the overview of the parameter selection.
All the identified parameters are system-dependent and vary based on the delays caused in the networking stack.
$\delta$ value is derived from the cycle time and offset.
If the \gls{cpu} is less powerful, the delays could get larger to accommodate the delays caused by the system before the packets reach the \gls{nic}.

\begin{table}
\centering
\caption{Identified optimum parameters}
\label{tab:opt-params}
\begin{tabular}{c c c c}
cycle time & \gls{cpu} isolation & offset & $\delta$    \\ \midrule
\SI{250}{\microsec} & on            & \SI{150}{\microsec}  & \SI{200}{\microsec}
\end{tabular}
\end{table}

During our experiments, we utilize \gls{uadp} over Ethernet. 
\Cref{fig:packetStructure} depicts the packet structure with the option to increase the payload in multiples of \SI{9}{\byte}.

\begin{figure}
\includegraphics[width=\columnwidth]{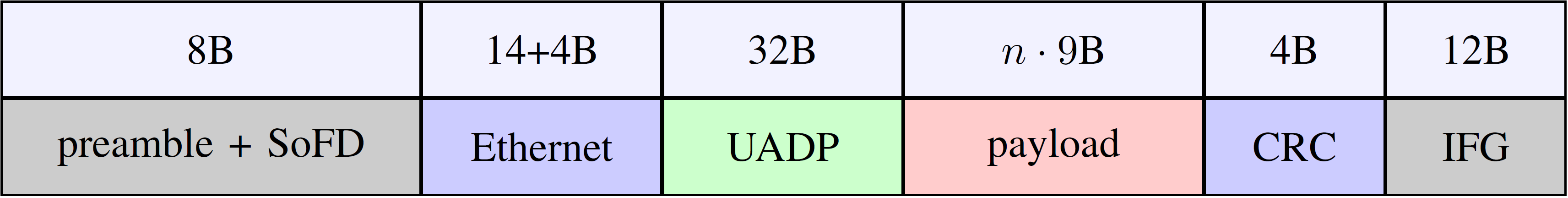}
\caption{Frame structure used in the experiments}
\label{fig:packetStructure}
\end{figure}

Our investigation into the impact of packet size on real-time performance is done by publishing \SI{8}{\byte} integers with \SI{1}{\byte} \gls{uadp} overhead in each packet.
The first test involves three variables, resulting in a frame size of \SI{81}{\byte}.
For subsequent tests, we use frame sizes closest to $80\cdot n$, where $n\in \{1,2,4,8,16,32\}$, and a maximum frame size of \SI{1521}{\byte}, which is close to the limit of VLAN-tagged Ethernet frames of \SI{1522}{\byte}. 
\Cref{tab:frame-sizes} shows the number of published variables, resulting sizes in different layers, as well as \gls{etf} results.
The drop rate remains below \SI{0.25}{\%} for frame sizes up to \SI{1278}{\byte}.
The publisher host experiences less than \SI{0.005}{\%} packet drops, even with maximum-sized frames.
However, the loopback host drops a quarter of all maximum-sized frames.
This suggests that the \gls{opc} \gls{ua} publisher can handle \SI{1521}{\byte} frames and is not the limiting factor.
The high drop rates observed in the loopback host may be attributed to longer traversal times and buffering in the networking stack after message reception.
Another possibility is that the time required for decoding and storing a received \gls{uadp} message exceeds a certain threshold.
New packets may enter the database at this threshold after the publisher reads from it.
In both scenarios, the drop rate comes from the oscillating behavior, where packets sometimes arrive on time and other times arrive too late.

\begin{table}
\centering
\caption{Mean results of ETF with increasing frame sizes}
\label{tab:frame-sizes}
\resizebox{\columnwidth}{!}{
\begin{tabular}{l c c c c c c}
\# variables & 3  & 12  & 30  & 65  & 136  & 163  \\ \toprule
\gls{uadp} [B]    & 59 & 140 & 302 & 617 & 1256 & 1499  \\
Link [B]    & \textbf{81} & \textbf{162} & \textbf{324} & \textbf{639} & \textbf{1278} & \textbf{1521} \\
Physical [B] & 101& 182 & 344 & 659 & 1298 & 1541  \\ \midrule
$d_P$ [\%]       & 0.0023    & 0.0027    & 0.0021    & 0.0029    & 0.0026    & 0.0045  \\
$d_L$ [\%]       & 0.0077    & 0.0503    & 0.0251    & 0.0603    & 0.245    &25.9 \\
$d_\Sigma$ [\%]  & 0.01    & 0.0529    & 0.0273    & 0.0631    & 0.247    &25.9  \\ \midrule 
RTT [µs]     & 500 & 500 & 500 & 501 & 501 & 570 \\
jitter [µs]  & 0.221 & 0.402 & 0.304 & 0.474 & 0.375 &136 \\ \bottomrule
\end{tabular}
}
\end{table}

A detailed analysis of the root cause of the poor performance observed with maximum-sized frames would require further experiments and measurement techniques.
Additionally, hardware and operating system must provide reliable real-time performance to exclude the impact of factors such as scheduling latencies.
However, the observation that all \gls{rtt} and jitter outliers occur at multiples of the cycle time indicates that the \gls{etf} \gls{qdisc} and Intel\textsuperscript{\textregistered} I210 \gls{nic} function as intended, sending packets at the scheduled time.
Thus, these components are not responsible for the poor performance.
Our results show that the maximum payload size that does not result in increased packet drops is \SI{1278}{\byte}.

We evaluate the \gls{cbs} \gls{qdisc} by testing different levels of over-provisioning and under-provisioning of $idleSlope_x$ values and measuring their impact on the performance. 
\Cref{tab:perf-p2p-cbs} shows that the drop rate is the lowest at an allocation of \SI{110}{\%}.
Our findings contrast with IEEE 802.1Qav, which suggests that \gls{cbs} should limit the flow bandwidth to the $idleSlope_x$.
The observed behavior is likely due to the \gls{cbs} \gls{qdisc} setting the credit to zero when the queue is empty.
When a burst of packets enters a \gls{cbs} queue, the first packet is sent when the credit is non-negative.
The credit then is decreased by the $sendSlope_x$ rate and may become negative.
If the credit is negative, the next packet in the queue must wait for the credit to increase.
This traffic pattern effectively limits the bandwidth to the $idleSlope_x$.
However, in our experiment, packets are sent every \SI{250}{\microsec}, leaving roughly a \SI{250}{\microsec} interval where no packet is enqueued.
This interval is too long for packets to queue up, resulting in only one packet being in the queue at a time.
Consequently, the credit is set back to zero shortly after sending a packet, allowing the next packet to be transmitted immediately.
The \gls{cbs} \gls{qdisc} of Linux may influence cyclic traffic with shorter cycle times and larger packets.
In such cases, the inter-frame gap becomes smaller, and packets start to queue up.

\begin{table}
\centering
\caption{Mean results of CBS with increasing $idleSlope_x$}
\label{tab:perf-p2p-cbs}
\begin{tabular}{l c c c c c}
$idleSlope_x$ [\%]           &  80 & 90 & 100 & 110  & 120  \\ \toprule
$d_P$ [\%]      & 0.0017    & 0.0017    & 0.0017    & 0.0016    & 0.0017   \\
$d_L$ [\%]      & 0.129    & 0.0948    & 0.137    & 0.0878    & 0.0948   \\
$d_\Sigma$ [\%] & 0.131    & 0.0965    & 0.139    & 0.0894    & 0.0965   \\ \midrule
RTT [µs]    & 250     & 252     & 251     & 251     & 251 \\
jitter [µs] & 1.2     & 2.13     & 0.911     & 0.68     & 1.1 \\ \bottomrule
\end{tabular}
\end{table}

For \gls{taprio}, we reserve one timeslot for \gls{opc} \gls{ua} traffic and the remaining portion of the cycle for best-effort traffic.
We gradually increase the priority timeslot to determine an appropriate duration by \SI{12.5}{\microsec}.
At the same time, we reduce the timeslot for best-effort traffic to achieve a cycle time of \SI{250}{\microsec}.
Furthermore, the priority traffic is protected by guard bands of \SI{15}{\microsec} before and after its timeslot to prevent interference from other traffic.
The guard band duration is chosen to be slightly longer than the serialization time of a maximum-sized 1GbE Ethernet frame.
According to \Cref{tab:perf-p2p-taprio}, all \glspl{ws} produce acceptable performance.
The mean jitter is \SI{0.2}{\microsec}, and \gls{rtt} is \SI{500}{\microsec}.
The drop rate decreases from \SI{0.0055}{\%} for \SI{12.5}{} and \SI{15}{\microsec} \gls{ws} to \SI{0.0035}{\%} for \SI{62.5}{} and \SI{75}{\microsec} \gls{ws}.
Therefore, a \gls{ws} of \SI{62.5}{\microsec} is used for the remainder of the experiments.

\begin{table}
\centering
\caption{Mean results of TAPRIO with increasing \gls{ws}}
\label{tab:perf-p2p-taprio}
\begin{tabular}{l c c c c c c}
\gls{ws} [µs]         &  12.5 & 25 & 37.5 & 50 & 62.5 & 75  \\ \toprule
$d_P$ [\%]       & 0.0011    & 0.0011    & 0.0005    & 0.0005    & 0.0001    & 0.0001   \\
$d_L$ [\%]       & 0.0044    & 0.0044    & 0.0038    & 0.0039    & 0.0034    & 0.0034   \\
$d_\Sigma$ [\%]  & 0.0055    & 0.0055    & 0.0043    & 0.0044    & 0.0035    & 0.0035   \\ \midrule
RTT [µs]     & 500    & 500     & 500      & 500     & 500     & 500 \\
jitter [µs]  & 0.21    & 0.21     & 0.21      & 0.208     & 0.208     & 0.209 \\ \bottomrule 
\end{tabular}
\end{table}

\subsection{Point-to-Point Topology}

We compare the \glspl{qdisc} in two \gls{p2p} topologies, E3 and D, equipped with different \glspl{cpu}.
\Cref{tab:perf-p2p-comparison-d} summarizes the results.
The drop rate of the publisher is lower than that of the loopback, and \gls{etf} on the publisher causes no packet drops, which validates the parameter selection.
However, both \gls{taprio} and \gls{etf} exhibit larger \gls{rtt} than other \glspl{qdisc} but show less jitter due to precise time control.
Additionally, setup D experiences fewer packet drops compared to setup E3.

\begin{table}
\centering
\caption{Comparison of setups E3 and D}
\label{tab:perf-p2p-comparison-d}
\begin{tabular}{l c c c c c}
\gls{qdisc} & \gls{etf} & \gls{fqcodel} & \gls{mqprio} &  \gls{cbs} & \gls{taprio} \\ \toprule
E3 $d_P$ [\%]      & 0.0023 & 0.0017 & 0.0017 & 0.0016  &N/A  \\ 
E3 $d_L$ [\%]       & 0.0077 & 0.0192 & 0.0939 & 0.0878 &N/A   \\
E3 $d_\Sigma$ [\%] & 0.01 & 0.0209 & 0.0956 & 0.0894 &N/A    \\
E3 RTT [µs]  & 500  & 250  & 251  & 251 &N/A     \\
E3 jitter [µs] & 0.221  & 0.428  & 0.999  & 0.680 &N/A     \\\midrule

D $d_P$ [\%]       & 0 & 0.0007 & 0.0007 & 0.0006 & 0.0001   \\ 
D $d_L$ [\%]     & 0.0033 & 0.0022 & 0.0026 & 0.0027 & 0.0034   \\
D $d_\Sigma$ [\%] & 0.0033 & 0.0031 & 0.0033 & 0.0033 & 0.0035   \\
D RTT [µs]    & 500  & 251  & 250  & 250  & 500    \\
D jitter [µs] & 0.21  & 13 & 8.92  &10.4  & 0.209    \\ \bottomrule

\end{tabular}
\end{table}

\Cref{fig:p2p-comparison-bartlisa-cdf} and \Cref{fig:p2p-comparison-rodtodd-cdf} depict the \gls{is} for setups E3 and D, respectively. 
\Cref{fig:p2p-comparison-bartlisa-cdf} shows an almost constant packet spacing of approximately \SI{250}{\microsec}, as the nearly straight line indicates.
However, the plot also shows significant outliers, which could render setup E3 unsuitable for specific real-time applications.
In contrast, \Cref{fig:p2p-comparison-rodtodd-cdf} shows that such outliers are absent for setup D, where the worst deviations are only ±\SI{60}{\microsec}.
Consequently, these findings suggest that setup D equipped with a more powerful \gls{cpu} may be better suited for real-time applications that require more bounded \gls{is}.
Overall, we see a trade-off between the standard deviation around the average and the worst-case performance.

\begin{figure*}
\centering
\begin{subfigure}{0.67\columnwidth}
\caption{P2P, host L, Setup E3}
\includegraphics[width=\columnwidth]{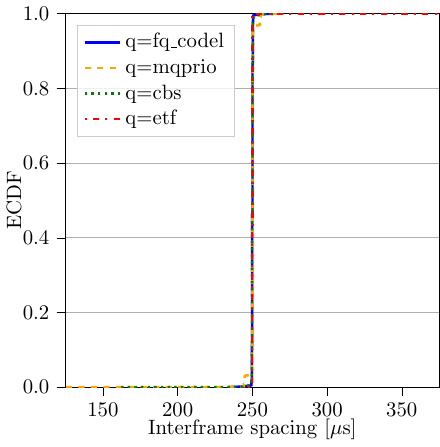}
\label{fig:p2p-comparison-bartlisa-cdf}
\end{subfigure}
\begin{subfigure}{0.67\columnwidth}
\caption{P2P, host L, Setup D}
\includegraphics[width=\columnwidth]{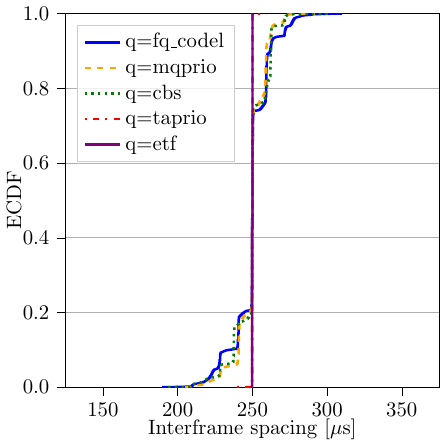}
\label{fig:p2p-comparison-rodtodd-cdf}
\end{subfigure}
\begin{subfigure}{0.67\columnwidth}
\caption{Bridged, host L, without BE traffic}
\includegraphics[width=\columnwidth]{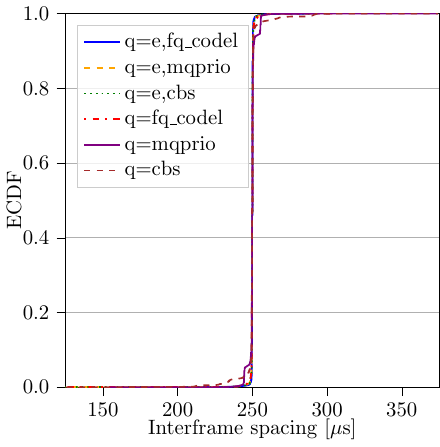}
\label{fig:bridge-comparison-cdf-lo-0M}
\end{subfigure}
\begin{subfigure}{0.67\columnwidth}
\caption{Bridged, host L, with 200 Mbps BE traffic}
\includegraphics[width=\columnwidth]{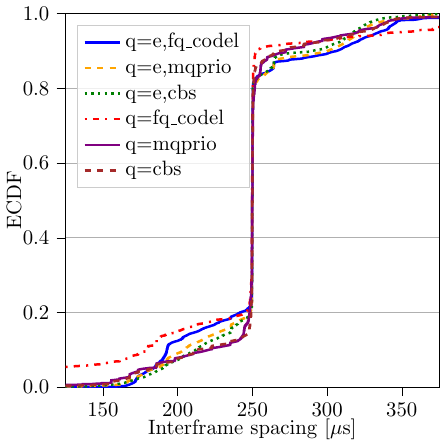}
\label{fig:bridge-comparison-cdf-lo-200M}
\end{subfigure}
\begin{subfigure}{0.67\columnwidth}
\caption{Bridged, bridge, without BE traffic}
\includegraphics[width=\columnwidth]{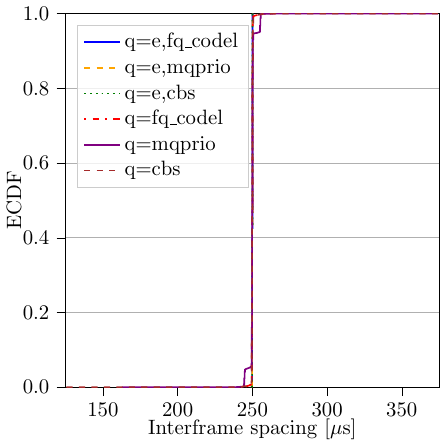}
\label{fig:bridge-comparison-cdf-bridge-0M}
\end{subfigure}
\begin{subfigure}{0.67\columnwidth}
\caption{Bridged, bridge, with 200 Mbps BE traffic}
\includegraphics[width=\columnwidth]{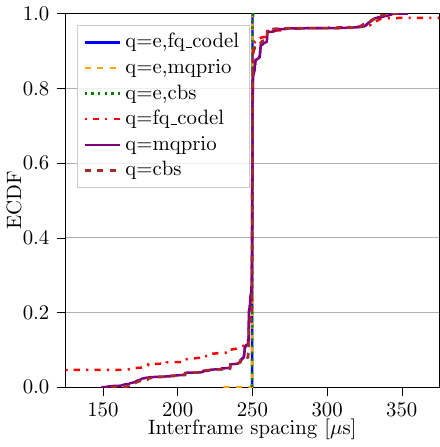}
\label{fig:bridge-comparison-cdf-bridge-200M}
\end{subfigure}
\caption{Comparison of \gls{is} of various \glspl{qdisc}}
\label{fig:p2p-comparison-qdiscs}
\end{figure*}

\subsection{Bridged Topology}

The \glspl{qdisc} are also compared in setup B, as shown in \Cref{fig:setup-b}, to investigate further the impact of an additional hop on the performance metrics, building upon the \gls{p2p} findings.
The additional hop is a Linux bridge.
We estimate the delay introduced by the bridge by comparing the \gls{rtt} in setup E3 and setup B.
However, a bug in the Linux networking stack prevents us from performing a parallel operation of \gls{ptp} with \gls{taprio} and \gls{etf}.
Thus, the results for this scenario are not available.
For the remaining \glspl{qdisc}, we use the parameters specified in \Cref{tab:opt-params}.
In addition, we introduce additional TCP \gls{be} cross traffic generated by the \textit{iperf3}\footnote{\url{https://iperf.fr/}} tool.
We observe a maximum rate of \SI{940}{\Mbps} between the end hosts for all \glspl{qdisc}, except when \gls{etf} is configured, which resulted in a lower rate of \SI{590}{\Mbps}.
It is worth noting that the \gls{be} traffic does not use any shapers but shares the physical interface.

\Cref{tab:perf-bridge-comparison} shows the results when no competing \gls{be} traffic exists.
The latency caused by the bridge is indicated as $l_B$. 
All \glspl{qdisc} achieve an acceptable drop rate without \gls{be} traffic.
The best overall drop rate of \SI{0.026}{\%} is achieved with \gls{etf} on the end hosts and \gls{mqprio} on the bridge.
The second-best results are obtained with \gls{fqcodel} on all hosts.
For all other \glspl{qdisc}, the drop rate is between \SI{0.14}{} and \SI{0.18}{\%}.
However, the less precise \gls{is} may cause the lower performance of \gls{cbs} and \gls{mqprio} without \gls{etf} at the loopback host, as shown in \Cref{fig:bridge-comparison-cdf-lo-0M}.
This explanation does not hold for the higher drop rate of \gls{etf} on the end hosts and \gls{fqcodel} or \gls{cbs} on the bridge, as it shows an almost perfect average behavior in \Cref{fig:bridge-comparison-cdf-bridge-0M}.
However, \gls{fqcodel} or \gls{cbs} on the bridge and \gls{etf} on the end hosts exhibit larger average and median \gls{rtt} than \gls{mqprio} on the bridge and \gls{etf} on the end hosts, leading to more packets missing their time slot for the loopback host's subscriber, resulting in packet drops.
When using \gls{be} cross-traffic, all \glspl{qdisc} exhibit an increased drop rate, as the processing overhead of the cross-traffic affects the other queues despite having a lower priority.

\begin{table}
\centering
\caption{Mean results on setup B without \gls{be} traffic}
\label{tab:perf-bridge-comparison}
\resizebox{\columnwidth}{!}{
\begin{tabular}{l c c c c c c}
\gls{qdisc}                     & E+FQ\_C & E+MQP & E+\gls{cbs} & FQ\_C     & MQP  & \gls{cbs}     \\ \toprule
$d_P$ [\%]               & 0.0047    & 0.0033  & 0.0018  & 0.0002    & 0  & 0 \\ 
$d_{b\rightarrow L}$ [\%] & 0    & 0.0002  & 0.0002  & 0.0007    & 0.0023  & 0.0007 \\
$d_L$ [\%]                & 0.171    & 0.022  & 0.135  & 0.054    & 0.149  & 0.15 \\
$d_{b\rightarrow P}$ [\%] & 0.001    & 0.0002  & 0.0002  & 0.0007    & 0.0007  & 0.0007 \\
$d_\Sigma$ [\%]           & 0.176    & 0.0257  & 0.137  & 0.0555    & 0.151  & 0.151  \\ \midrule
RTT [µs]             & 547     & 534   &  539  & 280     & 281   & 290   \\
$l_B$ [µs]           & 47.3      & 33.8    &  38.9   & 29.5      & 30.4    & 38.8   \\
jitter [µs]         & 2.31      & 1.06    & 1.12    & 0.84      & 2.25    & 1.40 \\ \bottomrule
\end{tabular}
}
\end{table}

To gain insight into the prioritization of the \glspl{qdisc}, we compare only the drop rates on the publisher host. 
\Cref{tab:perf-bridge-comparison-iperf} shows the drop rates, $d_P$, at different \gls{be} traffic rates.
Since \gls{etf} limits the bandwidth to approximately \SI{590}{\Mbps} on the end host, values for higher rates are represented as N/A.
The publisher host's performance is expected to be independent of the \gls{qdisc} configuration on the bridge.
However, we observe no clear drop rate trend for the publisher when using \gls{etf} on the hosts.
We also see no impact on the drop rate when increasing the \gls{be} traffic volume or using a different  \gls{qdisc} on the bridge.
As mentioned earlier, the \gls{etf} \gls{qdisc} drops packets that arrive after their TxTime, indicating a lack of resources.
Therefore, we assume that the packet misses the TxTime due to the limited resources, leading to packet drops.

\begin{table}
\centering
\caption{Publisher drop rates, $d_P$ [\%], on setup B with increasing \gls{be} traffic in [Mbit/s]}
\label{tab:perf-bridge-comparison-iperf}
\begin{tabular}{l c c c c c c}
\gls{qdisc}  & E+FQ\_C & E+MQP & E+\gls{cbs} & FQ\_C     & MQP  & \gls{cbs}     \\ \toprule
0    & 0.0047    & 0.0033  & 0.0018  & 0.0002    & 0  & 0 \\ 
200  & 0.0027    & 0.0023  & 0  & 0    & 0  & 0 \\ 
400  & 0.0017    & 0.0014  & 0.0128  & 0.0003    & 0  & 0 \\ 
600  & 0.0024    & 0.0038  & 0.0025  & 0    & 0  & 0 \\ 
800  & N/A          & N/A         & N/A         & 0.0003    & 0  & 0 \\ 
1000 & N/A          & N/A       & N/A        & 0.0028    & 0.012  & 0 \\
\bottomrule
\end{tabular}
\end{table}

Experiments conducted without \gls{etf} show a lower $d_P$.
The \gls{cbs} \gls{qdisc} on the publisher host does not drop any packet, regardless of the network load.
The same applies to \gls{mqprio} except for the congested network, i.e., at the rate of \SI{1000}{\Mbps}, where \gls{mqprio} drops \SI{0.012}{\%} of packets.
\gls{fqcodel} exhibits a packet drop rate between \SI{0}{} and \SI{0.0003}{\%}, corresponding to two packets being dropped among 700,000 sent packets.
In the congested network, \gls{fqcodel} drops \SI{0.0028}{\%}.
All \glspl{qdisc} effectively protect the higher priority traffic, with \gls{cbs} showing no packet drop and \gls{fqcodel} and \gls{mqprio} exhibiting no or very low packet drops, except for the congested network.
\gls{etf} shows the worst drop rate, but the numbers are biased due to the high worst-case scheduling latencies of the hosts in setup B.

The bridge introduces an additional latency overhead ranging from \SI{30}{} to \SI{47}{\microsec}, which we derive by subtracting the corresponding \gls{p2p} \gls{rtt} results from the \gls{rtt} results of setup B without \gls{be} traffic.
We observe that \gls{rtt}'s median and mean values in setup B without \gls{be} traffic are close, indicating that the outliers do not bias the results.
The lowest overhead of \SI{30}{\microsec} is observed for \gls{mqprio} and \gls{fqcodel} on all hosts.
In contrast, using \gls{fqcodel} only on the bridge and \gls{etf} on the end hosts causes the highest overhead of \SI{47}{\microsec}, indicating that \gls{fqcodel} works best when all hosts use the same \gls{qdisc}.
The overhead of \gls{cbs} compared to \gls{mqprio} is \SI{5}{} to \SI{8}{\microsec}.
The source of this overhead is unclear and may result from using \gls{cbs} together with \gls{mqprio}, which adds software component delay.
Further detailed analysis is required to investigate this overhead.

The average jitter is between \SI{0.8}{} and \SI{2.3}{\microsec}, with \gls{etf} setups on end hosts achieving smaller jitter, except for \gls{fqcodel}.
The bridge introduces a significant amount of jitter in the presence of \gls{be} traffic, but only a few when the network is idle.
This is illustrated with \gls{is} at the loopback host and the bridge for \SI{0}{} and \SI{200}{\Mbps} network load in \Cref{fig:bridge-comparison-cdf-bridge-0M} and \Cref{fig:bridge-comparison-cdf-bridge-200M}. 
\Cref{fig:bridge-comparison-cdf-bridge-200M} shows that the \gls{is} at the bridge is regular in the loaded network for \gls{etf} experiments on the end hosts.
After the packets traverse the bridge, they arrive at the loopback host, where more than \SI{20}{\%} of the packets have an \gls{is} smaller or larger than \SI{250}{\microsec}, as shown in \Cref{fig:bridge-comparison-cdf-lo-200M}.
This holds for all \gls{qdisc} configurations, including the \gls{etf} configured on end hosts.
Thus, the sending precision of these setups is lost once the packets leave the bridge.

All setups in the idle network, except for \gls{mqprio} on all hosts, achieve an almost regular \gls{is} at the bridge, as depicted in \Cref{fig:bridge-comparison-cdf-bridge-0M}.
However, after leaving the bridge, the \gls{ecdf} of \gls{is} shows slightly more variance for all experiments, except for \gls{cbs} on all setups.
The \gls{cbs} experiment shows the least regular \gls{is} after leaving the bridge, as shown in \Cref{fig:bridge-comparison-cdf-lo-0M}.

\subsection{Threats to Validity}
In this section, we discuss the limitations of our measurements and setups that may affect the validity of our results.
First, the \gls{ptp} time synchronization used in our experiments is imperfect.
While the hardware clocks are regularly adjusted and have a precision within tens of nanoseconds, the system and hardware clock synchronization is less precise, with the two clocks deviating up to \SI{2.7}{\microsec}.
Although this bias should be considered when synchronizing the application cycle to a \gls{taprio} cycle, we do not rely on software timestamps.
Hence, this offset is not considered in our measurements.
We believe the impact of clock deviations on our measurements is negligible, especially since the bias is small compared to the cycle time used in our experiments.
However, it is essential to note that we faced a challenge with \gls{ptp} and \gls{taprio}, which required dedicated wiring to synchronize clocks among different nodes \cite{10150312,9910175}.

Second, the expressiveness of some experiments may be limited by latency events caused by hardware and the operating system.
For instance, single outliers can significantly affect very low drop rates, but it can be challenging to detect such events.
We acknowledge that specialized hardware and software offering real-time performance can resolve this issue.
However, this approach may also limit the repeatability of experiments by the community.
To limit the impact of latency events, we explore real-time kernel possibilities for Linux and increase the number of samples and repetitions.
Additionally, we provide the same number of digits for performance parameters to improve the comparability of results in the tables.
However, for better readability, we round numbers in the text, which gives a correct impression of the precision of our measurements.

Overall, the results of our experiments using \gls{etf} or \gls{taprio} on all hosts offer a higher precision due to the deterministic sending of the hardware.
However, it is crucial to consider the limitations above when interpreting our results.
\section{Conclusion}
\label{sec:conclusion}

\balance

In conclusion, this study examines the real-time performance of \gls{opc} \gls{ua} regarding various \glspl{qdisc} and their configurations.
Drawing on an analysis of related works, we design and conduct experiments to bridge existing research gaps.
The results of the experiments are presented and evaluated.
The key findings of the study, which correspond to their respective research questions, can be summarized as follows.

\textbf{Q1, Q2:} The presented results demonstrate that synchronous \gls{tsn} scheduling \glspl{qdisc}, \gls{taprio} and \gls{etf}, are well-suited for periodic traffic with sub-millisecond cycle times due to their ability to control packet sending times precisely.
While the results fall within the use case requirements outlined in the introduction, further hardware and software optimization can reduce the number of outliers, thereby improving worst-case latency and jitter.
The remaining \glspl{qdisc}, \gls{mqprio}, \gls{cbs}, and \gls{fqcodel}, exhibit higher jitter but lower \gls{rtt} than \gls{taprio} and \gls{etf}, indicating they may be better suited for different use cases or scenarios.
Ultimately, the choice of \gls{qdisc} should be based on the application's specific requirements and the network conditions, considering the trade-offs between them.

\textbf{Q3:} The results of our investigation show that the Linux bridge introduces a two-way latency ranging between \SI{30}{} and \SI{47}{\microsec}.
Such latencies exist with competing \gls{be} traffic, dropping no more than \SI{0.004}{\%} of packets in the bridged setup while \glspl{qdisc} prioritize the \gls{opc} \gls{ua} traffic.
\gls{fqcodel} on all hosts achieves the best average \gls{rtt} overhead, while \gls{fqcodel} on the bridge and \gls{etf} on the end hosts cause the worst average overhead.
Our findings demonstrate that IEEE 802.1Qav \glspl{qdisc}, such as \gls{cbs} or \gls{mqprio}, are ineffective in limiting the drop rate.
However, \gls{fqcodel}, \gls{mqprio}, and \gls{cbs} introduce significant jitter in the presence of \gls{be} traffic, making them unsuitable for industrial real-time traffic in congested bridged networks.
Therefore, a one-hop setup without proper scheduling introduces enough nondeterminism for industrial real-time traffic to make the setup unsuitable.
Based on our results, any of the investigated non-\gls{tsn} \glspl{qdisc} may be used for the \gls{opc} \gls{ua} \gls{pubsub} application in a one-directional data transfer scenario, where higher jitter is not a concern.

\textbf{Q4:} For a better \gls{qdisc} performance, appropriate configuration based on the system's specification and traffic patterns is crucial.
For instance, \gls{mqprio} and \gls{taprio} require mapping of traffic classes to priorities, while \gls{cbs} requires four additional configuration parameters depending on the traffic pattern used.
\gls{taprio} also configures individual gate schedules and offset settings when used in TxTime-assisted mode.
Latencies between hosts can help optimize schedules, and setting the TxTime delay requires information about maximum latencies inside the hardware and software of a host.
\gls{etf}, which can be used as a child \gls{qdisc} of \gls{mqprio} or \gls{taprio}, only requires the configuration of a single value based on maximum latencies inside the host.
However, \gls{cbs} and \gls{etf} require at least an additional configuration effort of \gls{mqprio}.
Therefore, based on the ascending order of configuration effort, the ranking is \gls{fqcodel}, \gls{mqprio}, \gls{etf}, \gls{cbs}, and \gls{taprio}.
Following the presented guidelines, users can ensure optimum performance and traffic control with their chosen \glspl{qdisc}.

The appropriate \gls{qdisc} configuration is essential in achieving desirable performance, but it is not the only factor.
\gls{cpu} choice also impacts average and worst-case performance significantly.
Our results show that setup D, equipped with a powerful \gls{cpu}, outperforms setup E, equipped with a weaker \gls{cpu}, in the worst-case scenario.
Therefore, if real-time applications are the primary concern, a powerful \gls{cpu} is the better option due to its lower worst-case latency.
Considering \gls{qdisc} configuration and \gls{cpu} selection when designing and optimizing networked systems to achieve better performance is essential.

The results of our experiment demonstrate that the implementation of \gls{cbs} in Linux does not comply with the IEEE 802.1Qav standard.
Specifically, when the $idleSlope_x$ of a traffic class is set to \SI{80}{\%} of the bandwidth, \gls{cbs} is expected to drop \SI{20}{\%} of the packets.
However, our findings do not confirm that.
As a result, \gls{cbs} is ineffective in improving the \gls{opc} \gls{ua} \gls{pubsub} traffic performance with a cycle time of \SI{250}{\microsec} or more.
Based on these results, it is crucial to reconsider using \gls{cbs} in Linux for traffic management and explore alternative mechanisms that better align with the IEEE 802.1Qav standard.

In conclusion, the findings of this study emphasize the importance of carefully considering the requirements when deploying \gls{opc} \gls{ua} applications.
Our research demonstrates that \gls{cots} hardware and open-source software can effectively meet the real-time requirements of \gls{opc} \gls{ua} applications.
However, it should be noted that the performance of \glspl{qdisc} varies significantly depending on their configuration.
Therefore, it is recommended that system architects and engineers carefully evaluate the characteristics of the \glspl{qdisc} and configure them according to the specific requirements of the applications.
Overall, our research contributes to advancing the understanding of the optimum deployment of \gls{opc} \gls{ua} applications, and our findings aid in developing more efficient and effective \gls{opc} \gls{ua} systems.

\bibliographystyle{IEEEtran}
\bibliography{lit}

\end{document}